\documentclass[conference,twocolumn]{IEEEtran}
\usepackage{graphicx} 
\usepackage[margin=1in]{geometry}
\usepackage[colorinlistoftodos]{todonotes}
\usepackage{caption}
\usepackage{subcaption}
\usepackage{hyperref}
\usepackage{soul}

\usepackage{dsfont}
\usepackage{amsmath,amssymb,amsfonts, amsthm}
\usepackage{algorithmic}
\usepackage{textcomp}
\usepackage{bbm}
\PassOptionsToPackage{hyphens}{url}\usepackage{hyperref}
\usepackage{xcolor}
\def\BibTeX{{\rm B\kern-.05em{\sc i\kern-.025em b}\kern-.08em
    T\kern-.1667em\lower.7ex\hbox{E}\kern-.125emX}}

\usepackage[authoryear,round]{natbib}
\theoremstyle{plain}
\newtheorem{assumption}{Assumption}
\newtheorem{theorem}{Theorem}

\theoremstyle{definition}
\newtheorem{example}{Example}

\DeclareMathOperator{\E}{\mathbb{E}}
\DeclareMathOperator{\Var}{Var}
\DeclareMathOperator{\N}{\mathrm{N}}
\DeclareMathOperator{\Exp}{\mathrm{Exp}}

\DeclareMathOperator{\I}{\mathbb{I}}
\newcommand{\cdf}{F}

\newcommand*\diff{\mathop{}\!\mathrm{d}}

\usepackage{booktabs}
\begin{document}

\title{It's About Time: What A/B Test Metrics Estimate}

\author{\IEEEauthorblockN{Sebastian Ankargren}
\IEEEauthorblockA{\textit{Experimentation Platform} \\
\textit{Spotify}\\
Stockholm, Sweden \\
sebastiana@spotify.com}
\and
\IEEEauthorblockN{Mattias Frånberg} 
\IEEEauthorblockA{\textit{Experimentation Platform} \\
\textit{Spotify}\\
Stockholm, Sweden \\
mfranberg@spotify.com}
\and
\IEEEauthorblockN{Mårten Schultzberg} 
\IEEEauthorblockA{\textit{Experimentation Platform} \\
\textit{Spotify}\\
Stockholm, Sweden \\
mschultzberg@spotify.com}
}

\maketitle
\thispagestyle{plain}
\pagestyle{plain}
\begin{abstract}
Online controlled experiments, or A/B tests, are large-scale randomized trials in digital environments. This paper investigates the estimands of the difference-in-means estimator in these experiments, focusing on scenarios with repeated measurements on users. We compare cumulative metrics that use all post-exposure data for each user to windowed metrics that measure each user over a fixed time window. We analyze the estimands and highlight trade-offs between the two types of metrics.
Our findings reveal that while cumulative metrics eliminate the need for pre-defined measurement windows, they result in estimands that are more intricately tied to the experiment intake and runtime. This complexity can lead to counter-intuitive practical consequences, such as decreased statistical power with more observations. However, cumulative metrics offer earlier results and can quickly detect strong initial signals.
We conclude that neither metric type is universally superior. The optimal choice depends on the temporal profile of the treatment effect, the distribution of exposure, and the stopping time of the experiment. This research provides insights for experimenters to make informed decisions about how to define metrics based on their specific experimental contexts and objectives.
\end{abstract}

\section{Introduction}

\subsection{Contribution}

We study the properties of a common type of metric used to evaluate so-called A/B tests, which are randomized controlled trials used extensively by digital companies. The metric is known as a cumulative metric and uses all available data since receiving the treatment for each subject in the test. We compare the cumulative metric to a windowed metric, where each subject is only measured over a fixed window of time after they receive the treatment. Combining the two metrics results in a cumulative windowed metric, which measures each subject cumulatively but at most for a fixed amount of time.

In our analysis, we focus on the estimands of a difference-in-means comparison based on either of the three metrics to provide insight into what we can learn from them. We find that the estimand of the cumulative metric is a function of the underlying treatment effect curve as a function of time since exposure, the distribution of exposure, and the stopping time of the experiment. In contrast, the estimand of the windowed metric only depends on the window parameter and the underlying treatment effect curve. The estimand of the cumulative windowed metric is identical to the cumulative metric's initially, but converges to the windowed metric's.

The time-dependent estimand of the cumulative metric is not only of theoretical interest. We highlight the practical implications to show what the consequences are. 

First, the estimand's dependence on the distribution of exposure and the stopping time means that two experiments with the same underlying treatment effect curve answer different questions, unless the stopping time and the distribution of exposure are identical. As a result, the ability to compare and interpret estimated effects in A/B tests suffer.

Second, we analyze the statistical properties of tests based on these different metrics. We demonstrate that for cumulative metrics, the power of the test can paradoxically decrease with more data under certain conditions. 

Third, we explore the practical implications of these findings for experimental design, analysis, and decision-making in business contexts. We discuss how the choice of metric affects the interpretation of results, the comparability of experiments, and the alignment with business objectives. We also provide guidance on when cumulative metrics may still be appropriate, such as for error detection or in contexts with primarily temporary visitors.

\subsection{Estimands in scientific research and business contexts}

In scientific research, particularly in studies or experiments focused on causal effects, the estimand serves as a precise definition of the causal effect of interest \citep{rubin1974}. The estimand concept is fundamental to experimental design and analysis across various disciplines, including economics, medicine, and online experimentation. In many cases, a significant part of the study design involves identifying an estimator that can accurately measure the estimand of interest \citep{imbens2015causal}. 

A commonly used estimand is the Average Treatment Effect (ATE). Let $Y^W$ be the potential outcome under treatment $W\in \{0,1\}$. The ATE is defined as:

\begin{equation}
    \mathrm{ATE} = \E(Y^1 - Y^0).
\end{equation}

The wide use of the ATE as the main estimand is motivated by its suitability for decision-making purposes \citep{heckman2001policy}.

\subsubsection{Estimands in business contexts}

While the term \emph{estimand} is not commonly used in business settings, the concept is ubiquitous in corporate decision-making and performance measurement. Businesses frequently discuss and set precise targets for what they want to achieve. For example, a tech company might set goals for the growth of monthly active users as a key metric. When the company iterates to achieve this goal, they hope that the changes they make will cause a positive change for this metric. The estimand lets us make an equally precise definition of exactly what causal change we want to evaluate our efforts against. 

Precisely defined outcomes, or \emph{metrics}, that modern business use to track their goals go by many names. 
Objectives and Key Results (OKRs), a goal-setting framework popularized by companies like Google, define precise outcomes for evaluating business performance. The \emph{key results} of the OKR are quantifiable metrics that measure progress towards objectives \citep{niven2016objectives}. These metrics often represent causal effects of business strategies, analogous to treatment effects in scientific studies. Key Performance Indicators (KPIs) are another example of business estimands. They are measurable values that demonstrate how effectively a company meets key business objectives \citep{parmenter2015key}. Like scientific estimands, KPIs provide a precise definition of what is being measured and targeted. Similarly, Return on Investment (ROI) calculations in business contexts are conceptually similar to estimating treatment effects. They aim to quantify the causal impact of a business decision or investment on financial outcomes \citep{botchkarev2011return}. Lastly, Customer Lifetime Value (CLV) is a metric that businesses use to predict the total value a customer will bring to a company over their entire relationship. It involves complex calculations and predictions, much like estimating long-term treatment effects in scientific studies \citep{gupta2006modeling}. Arguably, the ATE of the CLV is the optimal estimand for experiments with its clear relevance for product and business decisions.

While the terminology differs between scientific and business contexts, there are clear parallels in definitions of precise, measurable outcomes and metrics. If precise and measurable outcomes are important for a company or organization, then so are the estimands targeted by their experiments.

\subsubsection{Experimental design requires estimands}
Estimands make up the basis for statistical inference, and therefore also for experimental design. In statistical theory, an estimator is considered unbiased if its expectation equals the true value of the estimand of interest. Multiple unbiased estimators can exist for the same estimand. For example, both the difference-in-means estimator and the regression estimator are unbiased for the ATE under randomized sampling and treatment design \citep{imbens2015causal}.

The properties of an estimator, together with hypothesized true values for the estimand, form the foundation for power analyses, which inform the experimenter on what sample size they need to collect. A fundamental assumption of power analyses is that the estimand and the estimator are fixed and prespecified. This usually means that the hypothesized true effect remains unchanged with the sample size, and that the variance decreases with the sample size. When this behavior no longer holds, power analyses behave counterintuitively and add much less value to the planning stage of an experiment. 

The notion that precision about certain parameters can decrease with more observations is not new in statistics. In the study of sequential testing, it is well known that the so-called statistical information \citep{fisher1925theory}---which is proportional to the inverse of the variance---for a parameter can decrease with more observations, see, e.g., \cite{shoben2017} for a recent example. The fact that the information is non-monotonically increasing for a parameter in a longitudinal setting can be more easily understood under the lens of estimands. Under certain data-generating processes, the estimand implied by an estimator changes, and new observations can both increase and decrease the information about the estimand.

In the context of online experiments, the most common estimand is the ATE with respect to a metric of interest. However, most experimental designs fail to make this explicit. A common consequence is that because the estimand is only loosely defined, changes to the experiment design that change the estimand often go unnoticed. We illustrate this in the next sections.

\section{Mathematical setup}

Consider an experiment that randomly distributes treatments $W_i\sim \mathrm{Bern}(p)$ to units $i=1, \dots, N$. We focus on users as our unit of randomization for readability, but the results generalize to other units. Users are exposed to the experiment at a random time $E_i>0$ after the experiment starts. We use the convention that $t=0$ denotes the start of the experiment. The experimenter measures an outcome of interest for each user. We denote a user's outcome at time $t$ by $Y_{i}(t)$. This measurement contains an appropriate summary of each individual's data on the interval $[e_i, t]$, where $e_i$ is the realized time of exposure for unit $i$. Our main focus in this paper are the implications of alternative ways to define $Y_{i}(t)$.

We next postulate the existence of two potential outcomes for each unit, which we denote by $Y^1_{i}(t)$ and $Y^0_{i}(t)$. The observed outcome for unit $i$ is $Y_{i}(t)=Y^{W_i}_{i}(t)$. Let  $R_{i,t}=\I(E_i<t)$ be an indicator for if user $i$ was exposed at time $t$, and $N_t=\sum_{i=1}^NR_{i,t}$ be the number of exposed users at time $t$. We let $N_t^1=\sum_{i=1}^NW_iR_{i,t}$ and $N_t^0=\sum_{i=1}^N(1-W_i)R_{i,t}$.

\begin{assumption}
We assume that: 
\begin{enumerate}
\item the Stable Unit Treatment Valuation Assumption (SUTVA) holds, so that there is no interference between users
\item there is perfect compliance, so that all users receive their assigned treatment
    \item the treatment assignment $W_i$ is independent of both the potential outcomes and the exposure time,
\begin{equation}
W \perp \left(Y^1(t), Y^0(t), E_i\right).
\end{equation}
\item the time of exposure $E$ is independent of both observed and unobserved characteristics $X$ and $U$,
\begin{align}
    E \perp (X, U).
\end{align}
\end{enumerate}
\label{assumptions}
\end{assumption}

These standard assumptions ensure that the experiment identifies the average treatment effect. To look at more specific estimands and how different measurement strategies affect them, we next make a simplifying assumption about the treatment effects.

\begin{assumption}
Each unit's outcome follows a stochastic process $\{Y(t), t \geq0\}$. The treatment effect at time $t$ given exposure at time $e<t$ is
\begin{multline}
\E[Y(t)|W=1, E=e]-\E[Y(t)|W=0, E=e]=\\\Delta(t-e), \quad e<t
\end{multline}
where 
\begin{align}
    \Delta(t)=\int_0^t\delta(s)\diff s.
\end{align}
\label{assumption}
\end{assumption}

\begin{figure}
     \centering
     \begin{subfigure}[b]{0.49\textwidth}
         \centering
         \includegraphics[width=\textwidth]{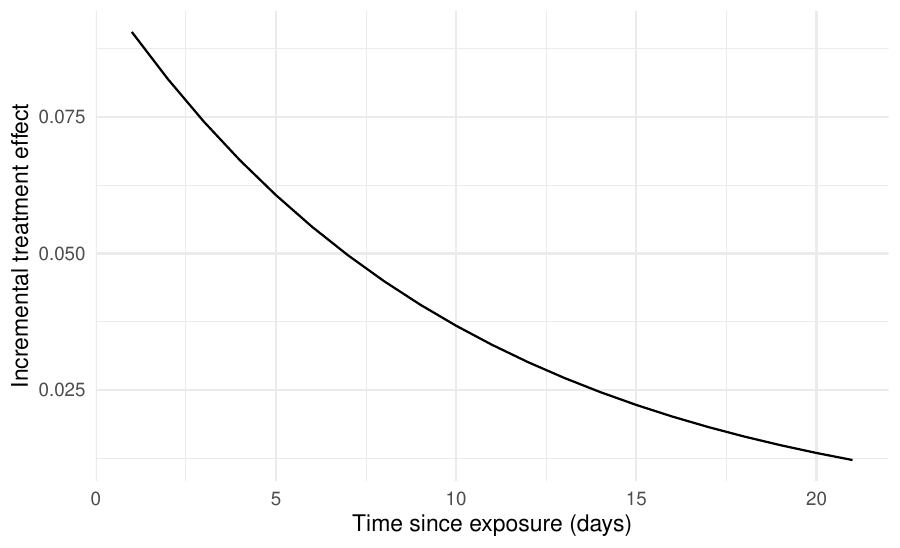}
         \caption{Incremental treatment effect $\delta(t)$}
     \end{subfigure}
     \hfill
     \begin{subfigure}[b]{0.49\textwidth}
         \centering
         \includegraphics[width=\textwidth]{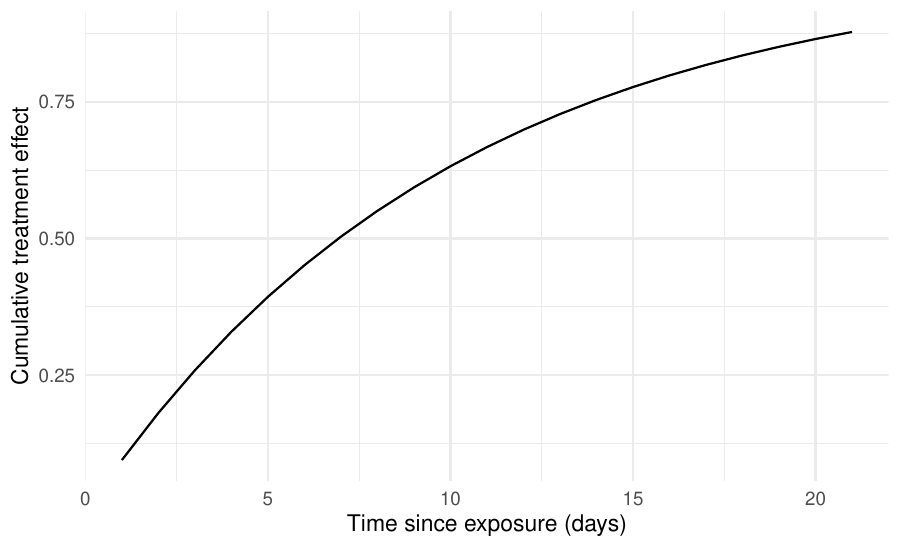}
         \caption{Cumulative treatment effect $\Delta(t)$}
     \end{subfigure}
        \caption{Examples of $\delta(t)$ and $\Delta(t)$. The curves describes the incremental and cumulative treatment effects at a given time as a function of the time since exposure. In the figures, $\delta(t)=0.1e^{-0.1 t}$}
        \label{fig:effects}
\end{figure}

The treatment effect curves, represented by $\delta(t)$ and $\Delta(t)$, play a central role in understanding the estimands of A/B test metrics. Figure \ref{fig:effects} shows an example of a large initial effect that decays with time since exposure. What the figure makes clear is that there is not \emph{one} treatment effect, but a treatment effect \emph{curve}. This stands in contrast to how most experimenters colloquially reason about treatment effects and metrics in A/B tests, but it underpins the single one-dimensional effects reported in the end. Ultimately, the new treatment experience will affect treated users differently over time. This means that the implied estimands are weighted sums of the treatment effect curves. As we show later in this section, the level of complexity in those weighted sums varies greatly. 

\subsection{Measurement strategies}
We now consider more specific measurement strategies that are functions of $\{Y_{i}(t)\}$. A \emph{metric} is a group-level average of the within-unit aggregations represented by $Y_{i}(t)$. For example, the metric \emph{Average minutes played per user} computes the average of $Y_{i}(t)$ for all users in a treatment group for a fixed $t$ according to
\begin{align}
    \hat{\mu}_t^{(1)}&=\frac{1}{N_t^{(1)}}\sum_{i=1}^N R_{i}(t)W_i Y_{i}(t)\\
    \hat{\mu}_t^{(0)}&=\frac{1}{N_t^{(0)}}\sum_{i=1}^N R_{i}(t)(1-W_i) Y_{i}(t).
\end{align}

In this case, $Y_{i}(t)$ represents the total minutes played between exposure and time $t$ for user $i$. The measurement is in itself an aggregation of more low-level data points. In the previous example, $Y_{i}(t)$ sums the length of each stream that occurred during the interval.

The specific form of $Y_{i}(t)$ used when computing the average influences the result. Importantly, depending on the exact form of $Y_{i}(t)$, the implied estimands differ. This means that A/B tests using metrics based on these different forms for $Y_{i}(t)$ answer different questions. As an example, consider an experimenter that instead wants to measure the metric \emph{Average minutes played per user during the first week after exposure}. This new metric includes a fixed time window over which each user should be measured. We next formalize three such variations, which are functions of $\{Y_{i}(t)\}$, and describe their estimands. 

In principle, all metrics defined below could be offset to start including measurements from each unit a certain number of periods after the unit was exposed. As all results in this paper are still relevant with exposure offsets, we leave this case out to simplify notation. 

\paragraph{Cumulative metrics}
Use $Y_{i}(t)$ without transformation from each unit. This means that units are measured on unequally long intervals, whose lengths equal $t-e_i$. By Assumption \ref{assumption}, the treatment effect conditional on exposure at time $e$ is
\begin{multline}
\E\left[Y(t)|W=1, E=e\right]-\E\left[Y(t)|W=0, E=e\right]\\
=\Delta(t-e).
\end{multline}

\paragraph{Windowed metrics}
Let $\nu>0$ be a window parameter fixed by the experimenter. Use
\begin{align}
Y^{\textrm{win}}(t)=
\begin{cases} 
\varnothing, \quad &\text{ if } e+\nu>t \\
Y(e+\nu), \quad &\text{ if } e+\nu\leq t
\end{cases}
\end{align}
for each unit. Units are measured over an interval of length $\nu$ after they are exposed. Their measurement is only made available after the full window $\nu$ has passed since their exposure. The treatment effect conditional on exposure at time $e$ is
\begin{multline}
    \E\left[Y^{\textrm{win}}(t)|W=1, E=e\right]\\
    -\E\left[Y^{\textrm{win}}(t)|W=0, E=e\right]\\=
    \begin{cases}
\varnothing, &\text{ if } e+\nu>t \\
\Delta(\nu), &\text{ if } e+\nu\leq t
\end{cases}
\end{multline}

\paragraph{Cumulative windowed metrics}
Let $\nu>0$ be the same window parameter used for windowed metrics. Use
\begin{align}
Y^{\textrm{cwin}}(t)=
\begin{cases} 
Y(t), \quad &\text{ if } e+\nu>t \\
Y(e+\nu), \quad &\text{ if } e+\nu\leq t
\end{cases}
\end{align}
for each unit. Similar to a windowed metric, the measurement is no longer updated after the full window $\nu$ has passed. The difference is that if the unit has been exposed for a shorter time than $\nu$, all of the available post-exposure data is used. The treatment effect conditional on exposure at time $e$ is
\begin{multline}
   \E\left[Y^{\textrm{cwin}}(t)|W=1, E=e\right]\\
   -\E\left[Y^{\textrm{cwin}}(t)|W=0, E=e\right]\\=
\begin{cases} 
\Delta(t-e),  &\text{ if } e+\nu>t \\
\Delta(\nu), &\text{ if } e+\nu\leq t
\end{cases} 
\end{multline}

\subsection{Estimands}
To derive the estimands at time $t$, we condition on exposure before time $t$. For the windowed metric, we condition on exposure before time $t-\nu$ so that the full window has passed and the estimand is defined.

\begin{theorem}
The estimands of the estimator applied to metrics based on the different measurement strategies are:
\begin{equation}
\begin{aligned}
    &\tau_C(t)\\
    &=\E\left[Y(t)|W=1, E<t\right]-\E\left[Y(t)|W=0, E<t\right]\\
    &=\frac{\int_0^t\Delta(t-e)\diff\cdf_E(e)}{\cdf_E(t)}
\end{aligned}
\label{eq:cumul}
\end{equation}

\begin{equation}
\begin{aligned}
    &\tau_W(t)\\
    &=\E\left[Y^{\textrm{win}}(t)|W=1, E<t-\nu\right]\\
    &-\E\left[Y^{\textrm{win}}(t)|W=0, E<t-\nu\right]\\
    &= \Delta(\nu)
\end{aligned}
\label{eq:window}
\end{equation}

\begin{equation}
\begin{aligned}
    &\tau_{CW}(t)\\
    &=\E\left[Y^{\textrm{cwin}}(t)|W=1, E<t-\nu\right]\\
    &-\E\left[Y^{\textrm{cwin}}(t)|W=0, E<t-\nu\right]\\
    &=\frac{\Delta(\nu)\cdf_E(t-\nu)+\int_{t-\nu}^t \Delta(t-e)\diff\cdf_E(e)}{\cdf_E(t)},
\end{aligned}
\label{eq:cumwin}
\end{equation}
where $\cdf_E$ is the cumulative distribution function of $E$.
\label{thm:thm1}
\end{theorem}

\begin{proof}
    See Appendix \ref{app:thm1}.
\end{proof}

The expressions for the estimands immediately reveal several characteristics of the different strategies. Focusing first on $\tau_C$ in \eqref{eq:cumul}, the estimand will vary with $t$ unless $\Delta(t)$ is a constant. This result implies that, at any two different points in time, the estimand of a difference-in-means comparison for a cumulative metric is not the same. For a given treatment effect curve $\delta$ and a fixed time since exposure $t$, the estimands of two different experiments will differ unless the distribution of exposure is the same. Intuitively, the estimand is the (scaled) area under the cumulative treatment effect curve weighted by the share of users exposed for different lengths of time.

The estimand for the windowed metric, denoted by $\tau_W$ in \eqref{eq:window}, is a much simpler expression: the cumulative treatment effect measured at the end of the window $\nu$. This result shows that the distribution of exposure does not change the estimand, nor by the time $t$ of the analysis. The window parameter $\nu$ affects the estimand, but the advantage is that the experimenter makes a clear and transparent choice. The downside is that the estimand is only defined when $t>\nu$. This means that there is no well-defined estimand in the beginning of an experiment. For an experiment with daily data and $\nu=7$, no effect exists during the first week.

The estimand from the cumulative windowed metric in \eqref{eq:cumwin} is combination of the previous two estimands. An alternative way to express it is as
\begin{align}
    \tau_{CW}(\nu, t)=\tau_W(\nu)c_1(t)+[\tau_C(t)-\tau_C(t-\nu)]c_2(t).
\end{align}
The weight $c_1(t)=0$ if $t-\nu<0$ and $c_1(t)\to 1$ as $t\to\infty$. At the same time, $\tau_C(t-\nu)=0$ if $t-\nu<0$, and $[\tau_C(t)-\tau_C(t-\nu)]c_2(t)\to 0$ as $t\to\infty$. This showcases how $\tau_{CW}$ initially is equal to $\tau_C$, and then converges to $\tau_W$. These relations only hold when the windowed and cumulative windowed metric measure users immediately after exposure, while exposure offsets that delay when to start measuring are common in practice. However, these relations between the estimands are helpful for building intuition about how the metrics differ more generally.

To more easily interpret the estimands, we next illustrate them with examples.

\begin{example}

\begin{figure}
     \centering
     \begin{subfigure}[b]{0.49\textwidth}
         \centering
         \includegraphics[width=\textwidth]{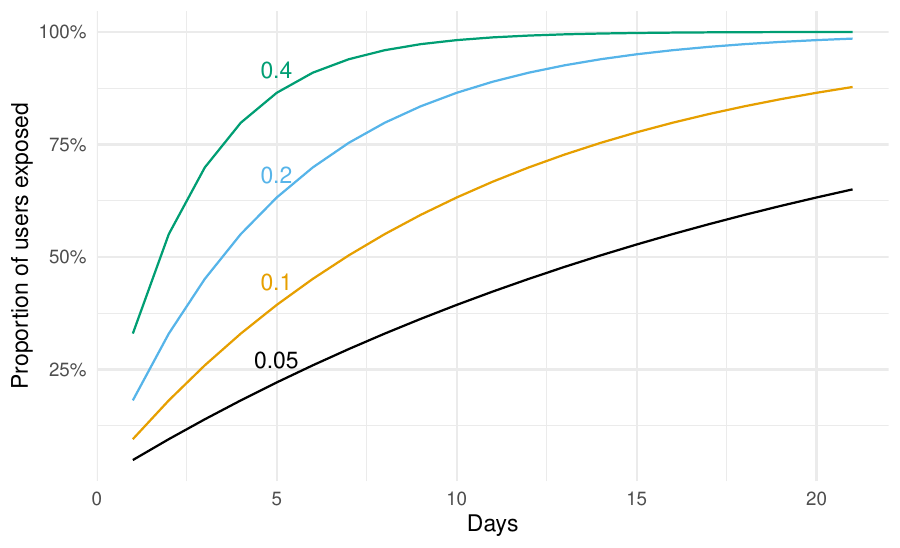}
         \caption{Exposure curves. Each curve shows the cdf of an $\mathrm{Exp}(\lambda)$ distribution.}
     \end{subfigure}
     \hfill
     \begin{subfigure}[b]{0.49\textwidth}
         \centering
         \includegraphics[width=\textwidth]{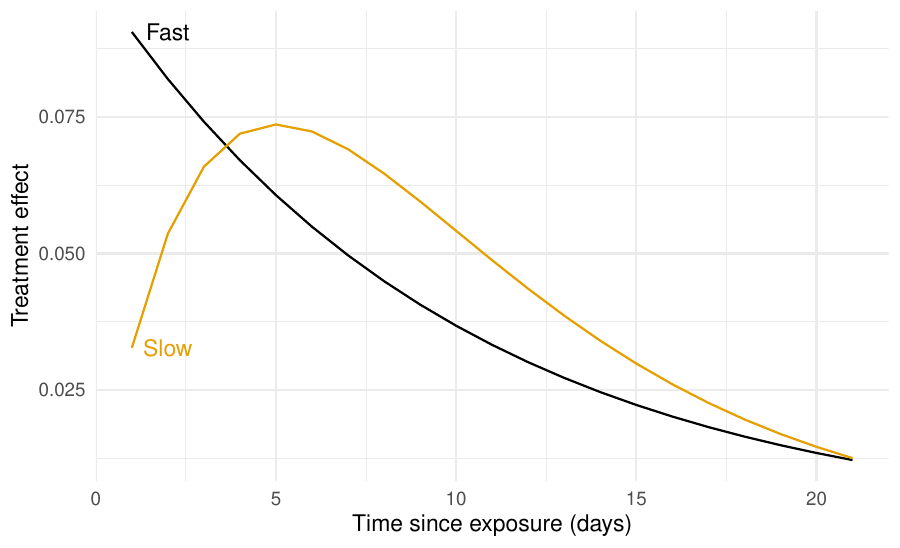}
         \caption{Effect curves. Each curve exemplifies the treatment effect curve as a function of days since exposure.}
     \end{subfigure}
        \caption{Examples of exposure and treatment effect curves.}
        \label{fig:example-1}
\end{figure}

We first assume that the time of exposure is exponentially distributed, $E_i\sim \mathrm{Exp}(\lambda)$. In practice, an exponential distribution often approximates the distribution of exposed users well. For the treatment effect $\delta(t)$, we consider two effect curves. The first effect curve is $\delta(t)=0.1e^{-0.1t}$ and decays exponentially from the point of exposure, illustrating a fast response to the treatment. The second curve is $\delta(t)=0.04te^{-0.2t}$ and peaks on day 5 after exposure, showing a slower response to the treatment. Figure \ref{fig:example-1} shows what these curves look like over the first 21 days. We set the window parameter $\nu=7$ to mimic an interest in effects during the first week after exposure.

\begin{figure}
     \centering
     \begin{subfigure}[b]{0.49\textwidth}
         \centering
         \includegraphics[width=\textwidth]{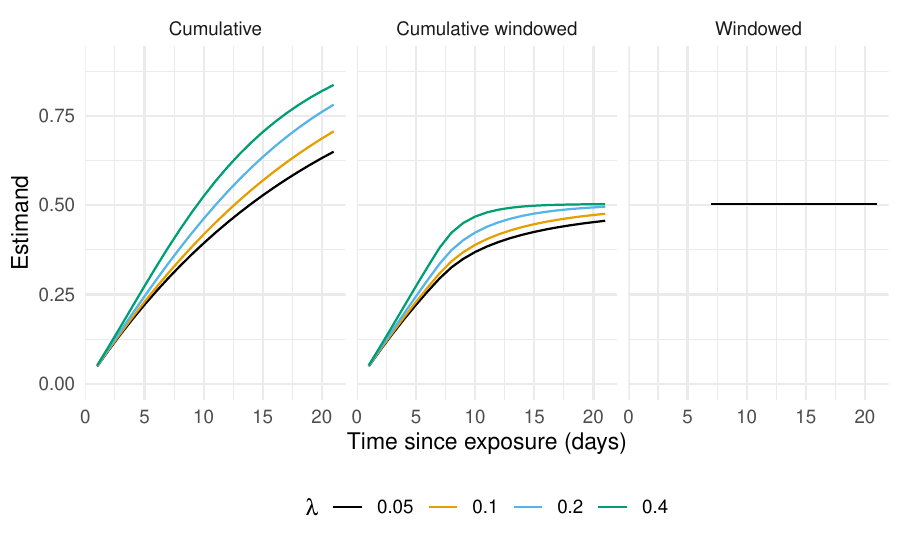}
         \caption{Estimands with $\delta(t)=\delta_{\text{fast}}(t)$.}
     \end{subfigure}
     \hfill
     \begin{subfigure}[b]{0.49\textwidth}
         \centering
         \includegraphics[width=\textwidth]{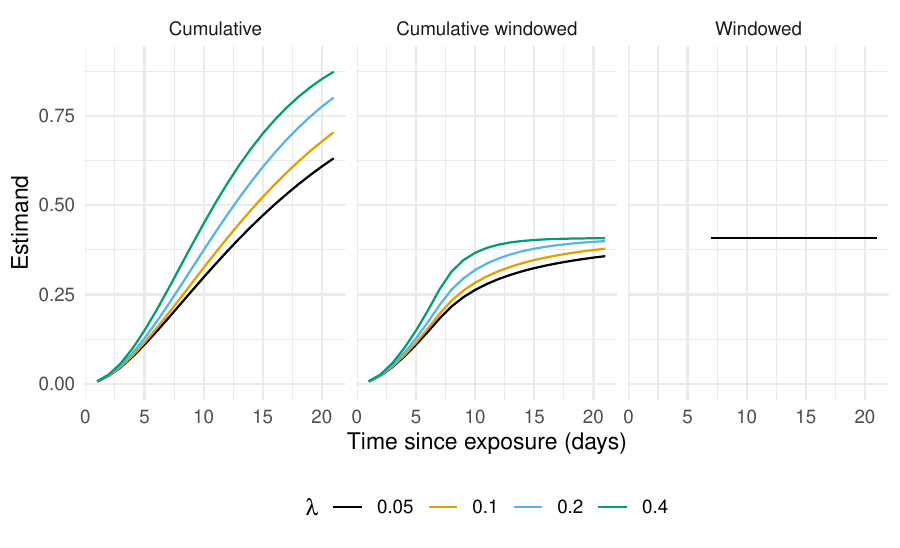}
         \caption{Estimands with $\delta(t)=\delta_{\text{slow}}(t)$.}
     \end{subfigure}
        \caption{Estimands. Exposure is exponentially distributed $E\sim \mathrm{Exp}(\lambda)$ and treatment effect curves are $\delta_{\text{fast}}(t)=0.1e^{-0.1t}$ and $\delta_{\text{slow}}(t)=0.04te^{-0.2t}$, see Figure \ref{fig:example-1}. }
        \label{fig:example-1-est}
\end{figure}

The left panel of Figure \ref{fig:example-1-est} shows the estimands as a function of time since exposure, and the parameter $\lambda$ that controls the rate at which users are exposed. The most striking result is the difference in behavior between the cumulative and windowed metrics. The two cumulative metrics change as a function of the time since exposure. Additionally, their estimands differ on a given day as a function of the rate of exposure $\lambda$. These results mean that, with a cumulative metric, two experiments that have the exact same distribution of exposure and treatment effects, will not estimate the same estimand unless they run for the same number of days. 

\end{example}

\subsection{Relation to previous research}
\cite{Callaway2021} study estimands in the presence of so-called staggered adoption when using difference-in-differences for observational data. Working in a discrete time setting, their key building block is the group-time average treatment effect parameter $\mathrm{ATT}(e,t)$. In our setting and with slightly adjusted notation,
\begin{equation}
\begin{aligned}
    \mathrm{ATT}(e,t)=\Delta(t-e).
\end{aligned}
\end{equation}

To summarize group-time average treatment effects, \cite{Callaway2021} use
\begin{align}
    \theta_t=\sum_{e\in\mathcal{E}_t}\sum_{s=0}^tw(e,s,t)\times \mathrm{ATT}(e,s),
\end{align}
where $\mathcal{E}_t$ is the set of all time points less than or equal to time $t$ at which users have been exposed. For the weights $w(e,s,t)$, \cite{Callaway2021} write that they are "carefully-chosen (known or estimable) weighting functions specified by the researcher such that $\theta$ can be used to address a well-posed empirical/policy question". If the parameter of interest is the cumulative metric estimand $\tau_C(t)$, then Theorem \ref{thm:thm1} reveals that the $w(e,s,t)$ weight is
\begin{align}
    w(e,s,t)=\frac{\Pr(E=e)}{\cdf_E(s)}\I(s=t).
\end{align}
There is nothing carefully chosen about this weight. The distribution of exposure---and not the researcher---determines the weights, and, by extension, the empirical question addressed.

In randomized experiments, the presence of treatment effects that vary with the time since exposure has received some attention in stepped-wedge cluster randomized trials. \cite{Hughes2015} call this \emph{time-on-treatment effects} and discuss how to use model-based estimation to make inference for these effects. \cite{Maleyeff2023} discuss various estimands of interest and model-based inference for the treatment effect curve, while \cite{Lee2024} investigate how misspecification of the treatment effect curve in mixed effects models affects inference. Most closely related to our work, \cite{Kenny2022} describe various estimands of interest. These include the \emph{time-averaged treatment effect (TATE)} obtained as $t^{-1}\Delta(t)$, or the \emph{long-term treatment effect (LTE)}, defined as $\delta(t)$ for a large enough $t$, which conceptually is meant to capture the treatment effect after it has flattened out. 

Our focus in this paper is on understanding what commonly used approaches to metrics in A/B tests estimate, whereby we rely on non-parametric estimation of treatment effects via difference-in-means.  

\section{More data is not always better}\label{sec:power}
We next investigate the power of cumulative metrics, with power meaning the probability for the test to reject the false null hypothesis of no effect in favor of the alternative hypothesis that there is some effect. We begin by showing a result for cumulative metrics that states that power can decrease with more data.

To this end, consider a comparison of means based on cumulative metrics, whose $Z$ statistic is equal to
\begin{align}
    Z_t = \frac{\hat{\mu}_t^{(1)} - \hat{\mu}_t^{(0)}}{V_t},
\end{align}
where $V_t=\Var \left(\hat{\mu}_t^{(1)}|\mathcal{R}_t\right)+\Var\left(\hat{\mu}_t^{(0)}|\mathcal{R}_t\right)$ is the (known) variance of the mean difference and $\mathcal{R}_t=\{R_1(t), \dots, R_N(t)\}$.

\begin{theorem}\label{thm:power}
    Consider a one-sided test for a difference in means based on the $Z$ statistic $Z_t$. The test rejects the null hypothesis of no difference if $Z_t>c$. Suppose the true effects are given by the sequence $\{\tau_C^1(t)\}$, that $\hat{\mu}_t^{(1)}$ and $\hat{\mu}_t^{(0)}$ are both normally distributed, and that Assumption \ref{assumptions}-\ref{assumption} hold. Then the power of the test at a later time point $t'$ is lower than the power at the earlier time point $t$ if
\begin{equation}
\begin{aligned}
&V_{t'}^{-1/2}\tau_C^1(t')-V_t^{-1/2}\tau_C^1(t)\\
    &=\left[\frac{1}{V_{t'}^{1/2}\cdf_E(t')}-\frac{1}{V_{t}^{1/2}\cdf_E(t)}\right]\int_0^t\Delta(t-e)\diff\cdf_E(e)\\
&+\frac{1}{V_{t'}^{1/2}\cdf_E(t')}\int_0^t\int_{t-e}^{t'-e}\delta(s)\diff s\diff\cdf_E(e)\\
&+\frac{1}{V_{t'}^{1/2}\cdf_E(t')}\int_t^{t'}\Delta(t'-e)\diff\cdf_E(e)\\
&<0.
\end{aligned}
\label{eq:diff}
\end{equation}
\end{theorem}
\begin{proof}
    The distributions of the statistics are $Z_t\sim \N\left(V_t^{-1/2}\tau_C^1(t), 1\right)$ and $Z_{t'}\sim \N\left(V_{t'}^{-1/2}\tau_C^1(t'), 1\right)$. If $V_{t'}^{-1/2}\tau_C^1(t')<V_t^{-1/2}\tau_C^1(t)$, then $Z_t$ stochastically dominates $Z_{t'}$ in the first order and $1-F_{Z_t}(z)>1-F_{Z_{t'}}(z)$ for all $z\in\mathbb{R}$ \citep{Petrova2018}. See Appendix \ref{app:thm2} for a derivation of the decomposition in the expression. .
\end{proof}
Theorem \ref{thm:power} specifies the conditions under which power decreases with more observations for the $Z$ test under normality.

The expression in Equation \eqref{eq:diff} shows how the difference in the expected $Z$ statistic between two time points consists of three parts. The first term captures the change in the contribution from effects through time $t$ from users that were exposed prior to time $t$. The first factor in the first term adjusts for the change in variance between the two time points. Because of the cumulative measurements, $V_t$ changes both because of more users and because of more data for existing users. The second term describes the change from treatment effects on the interval $(t, t')$ from users exposed before time $t$. The final term describes how users exposed between $t$ and $t+1$ contribute to the difference.

Power is lower at time $t'$ than at time $t$ if the difference in Equation \eqref{eq:diff} is negative. As the expression reveals, whether this occurs is a complicated function of the exposure distribution, the cumulative treatment effects, and the variance. New data can increase the variance because users are measured for longer periods, which makes the first term negative. If the exposure distribution gives large weight to periods with weaker treatment effects, the second term is only marginally positive. The third term can be exactly zero if no new users are exposed between $t$ and $t'$. It can be small if only few new users are exposed. Whether the overall difference is positive or negative thus depends on the exact nature of the treatment effects and the exposure distribution. A negative difference, and a reduction in power from more data, is more likely to happen when the cumulative treatment effect curve $\Delta$ flattens out over time and at times after most users have been exposed. In these cases, new data dilutes existing data with data from periods where treatment effects are lower. We illustrate this more concretely in an example.

\begin{example}
\label{ex:ex1}
Consider an experiment with 1,000 users equally distributed into a treatment and a control group. The experiment exposes 500 users at $t=0$ and the remaining 500 at $t=7$. The incremental treatment effect is $\delta(t)=c$ for $t\in(0,7)$ and 0 otherwise, which means that the cumulative treatment effect is $\Delta(t)=tc$ for $t\in(0, 7)$ and $7c$ for $t>7$. Suppose that the variance of $Y(t)$ grows linearly and is equal to $\Var[Y(t)|E=e]=\sigma^2(t-e)$. This assumption includes common stochastic processes like Brownian motions and compound Poisson processes. Then the expectation of the $Z$ statistic is
\begin{align}
        \E(Z_t)=\begin{cases}
        \frac{125c}{\sigma^2}, &\quad t< 7\\
        \frac{500tc}{4\sigma^2\left(t-3.5\right)+c^2\left(t-14\right)^2}, &\quad 7\leq t < 14\\
        \frac{250\times 7c}{\sigma^2(t-3.5)}, &\quad t\geq 14
        \end{cases}
\end{align}
The expectation represents the location of the sampling distribution under the alternative hypothesis, and that location is in this example either constant or decreasing in $t$. Figure \ref{fig:ex1} shows $\E(Z_t)$ with $c=0.05$ and $\sigma^2=2$. See Appendix \ref{app:ex2} for details.

\begin{figure}
\centering
    \includegraphics[width=0.5\textwidth]{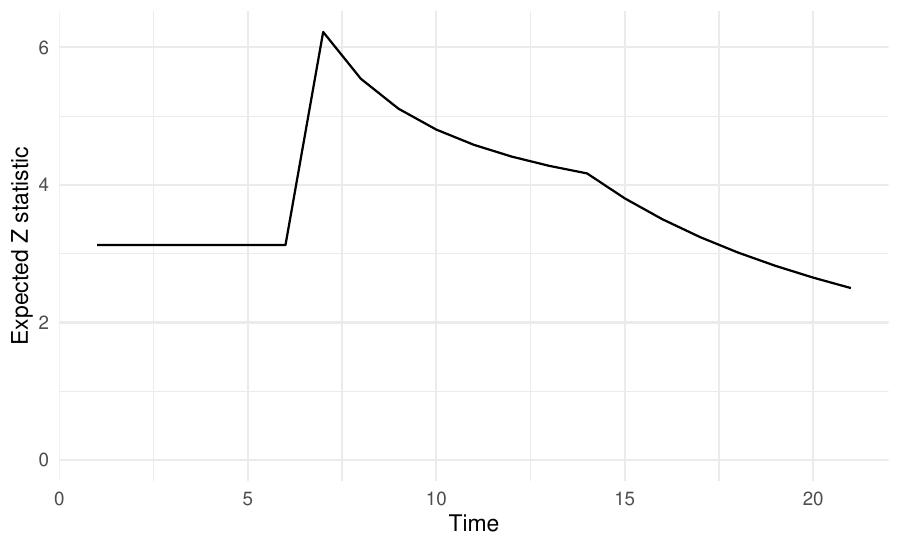}
    \caption{$\E(Z_t)$ for Example \ref{ex:ex1} with $c=0.05$ and $\sigma^2=2$.}
    \label{fig:ex1}
\end{figure}

The expectation is initially flat because the difference in means and its variance is affected equally by more data. At $t=7$, a new batch of users is exposed, which contributes with a stronger signal of an existing treatment effect. However, from $t=7$ and onwards, any new data for users that were exposed at $t=0$ contains no new information about a treatment effect since $\delta$ is 0 from this point on. The new data inflates the variance, which causes $\E(Z_t)$ to decrease. Starting at $t=14$, all data carrying information on a treatment effect has already been consumed, and any additional data only inflates the variance and dilutes the effect.

\end{example}

\subsection{Simulated power examples}
To further illustrate the non-intuitive ways that the power for cumulative metrics depend on the exposure and effect curves, we conduct two Monte Carlo simulations. These simulations are not intended to prove the superiority of any type of metric. Instead, they are meant to give further illustrations of the power behaviour of the three kinds of metrics. 

\subsubsection{Setup}
We generate data from independent and identically distributed standard normal distributions $\N(0,1)$ between and within units. We generate data for 700 units for each replication, and split the users into two equally sized groups. The maximum length of the experiment is 21 days. The number of time points included in the analysis per user depends on when the user was exposed. For each data-generating process (DGP), we repeat the simulation for 10,000 replications.
We compute the power of the three metrics considered earlier: a cumulative metric, a cumulative windowed metric with a window of 7 days, and a windowed metric with a window of 7 days. The simulation estimates the empirical power by calculating the rate of replications that reject the null hypothesis of no mean difference at time $t$. The false positive rate used in the test is 10\%.

We consider two different exposure distributions and treatment effect curves, described below.

\subsubsection{Data-generating process I}
For the first case, we let the exposure distribution be $E\sim \Exp(0.4)$ distribution, and the effect curve $\delta$ be $\delta(t)=e^{-t}$. In this case, the effect decreases rapidly after the point of exposure, and the exposure distribution decreases quickly after the start of the experiment. 

\begin{figure}
    \centering
    \includegraphics[width=0.5\textwidth]{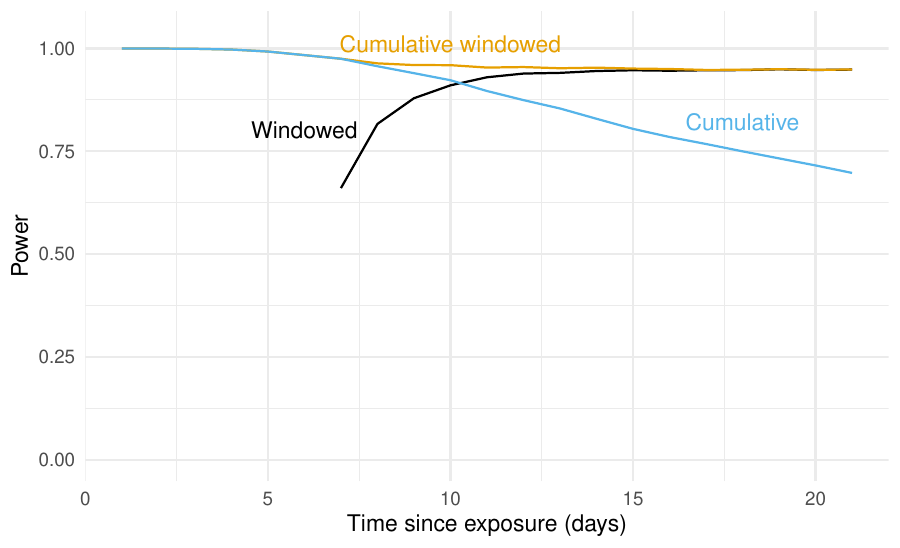}
    \caption{Power for cumulative, cumulative windowed, and windowed metrics at different stopping times. The exposure distribution is $E\sim \Exp(0.4)$ and the treatment effect curve $\delta$ is $\delta(t)=\delta^{-t}$.}
    \label{fig:exp(0.4)}
\end{figure}

Figure \ref{fig:exp(0.4)} displays the results. 
As expected, the power of the cumulative metrics is high early in the experiment, since all exposed units have only been measured for time points where the effect is large. As the experiment progresses, the share of users that have only been measured for time points with large effects decreases. At the same time, units with observations for which the true effect is small increases, which drives power down. This decline in power only has a limited effect on the cumulative windowed metric, since it does not include measurements for users beyond the seventh day. The windowed metric displays result first on the seventh day after experiment start, at which point its power is lower than the cumulative windowed metric's.  Further into the experiment, more users have been exposed for at least seven days and the two windowed metrics converge. The cumulative metric, however, keeps including results from later days for an increasing amount of users. The decline in new users that are exposed to the experiment fails to make up for the daily dilution of the effects, which leads to a steady decline in power.

\subsubsection{Data-generating process II}
In our second data-generating process, we let the exposure distribution be proportional to $t^3$, and let the treatment effect curve $\delta$ be $\delta(t)=\frac{14^{84}}{\Gamma(84)}t^{83}e^{-14t}$. In this case, the treatment effect peaks around day 6 after exposure and is close to zero at all other time points. The rate at which users are exposed increases throughout the experiment.

\begin{figure}
    \centering
    \includegraphics[width=0.5\textwidth]{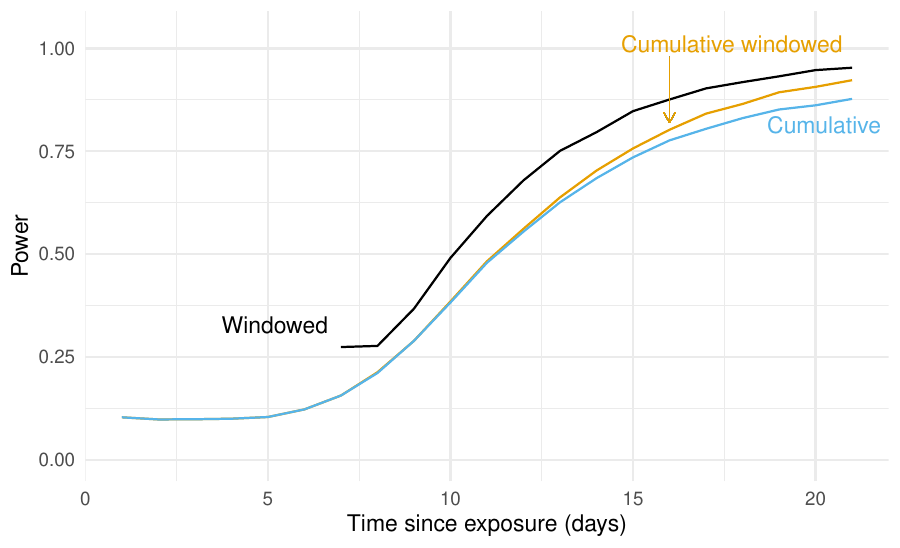}
    \caption{Power for cumulative, cumulative windowed, and windowed metrics at different stopping times. The exposure distribution is proportional to $t^3$ and the treatment effect curve $\delta$ is $\delta(t)=\delta(t)=\frac{14^{84}}{\Gamma(84)}t^{83}e^{-14t}$.}
    \label{fig:cubed}
\end{figure}

Figure \ref{fig:cubed} displays the results.
The power for both of the cumulative metrics hovers around the false positive rate for the first days, since the effect is close to zero for the first five days after exposure. After day five, the power increases at a similar rate for both cumulative metrics. Towards the end of the experiment, the cumulative windowed metric has slightly higher power, which is expected as the effect per unit is less diluted for the cumulative windowed metric than the cumulative metric. The windowed metric has higher power from its first result and onward since it never includes any users without a treatment effect. Moreover, the power of the cumulative windowed metric and the windowed metric have not yet converged in this case. Because the rate of exposure increases, new users enter the experiment every day of the experiment. For their first days, the effect is negligible and thus constantly pulls the power of the cumulative windowed metric down.

\section{Implications}
\subsection{Policy and decision making}
The results in Figure \ref{fig:example-1-est} and Equation \eqref{eq:cumul}-\eqref{eq:cumwin} establish that the estimands of a metric that does not include a window is highly dependent on aspects of the experiment design that are only partially, at best, under the control of the experimenter.

The changing estimand not only affects the interpretation of experiment results in relation to business metrics like OKRs and KPIs. The comparisons of results between experiments are also affected. Two experiments that run for different lengths will rarely be exactly comparable. 
With cumulative metrics, even two identical experiments that test the same change using the same outcome can produce very different answers depending on the length of the experiment itself. For windowed metrics, the experimenter is explicit about what they are interested in measuring, which makes reasoning about the treatment effect substantially less complex.   

For a company or organization for which the magnitudes of treatment effects matter, an ill-defined estimand violates a principle of clearly defined outcomes. Many companies and organizations work with outcome metrics that are precisely defined, and likely expect experimenters to present the impact of changes and innovation in terms of equally well defined estimands, if not exactly the same ones. However, not all experiments are equal and we return to when to use cumulative metrics in Section \ref{sec:when_to_use_cum_metrics}.

\subsection{Designing and planning experiments}
The results in Section \ref{sec:power} show that the power to reject the null under certain true treatment effects behaves in a non-intuitive way. Meanwhile, high quality experiments rely on power analysis. Modern experimentation tools take many aspects into consideration for calculating the sample size required to control the false positive and negative risks below rates decided by the experimenter \citep{schultzberg2024risk}. Importantly, all power analysis is based on the assumption of a fixed estimator that the variation in the sampling distribution can be estimated for. 

A fundamental principle for power analysis in online experiments is that the variance of the measurement in the metric is assumed fixed in the target population, so that the variance of the estimator is a monotonically decreasing function of the sample size. Because the measurements, and hence their variances, change with the runtime of the experiment, there is not one single variance to estimate for a cumulative metric. This means that the power analysis can no longer accept a fixed variance as the input and output the required sample size, because the variance intrinsically depends on the sample size across both users and time.

The power analysis challenges remain during the experiment. Many experimenters use sequential or fixed-power designs \citep{nordin2024precision}, and monitor confidence interval widths and current required sample sizes to learn whether the experiment will reach the intended level of precision in an acceptable time frame. For example, on the day the experiment is planned to end, the experimenter sees that the experiment has almost collected enough users, and decides to keep it running for another day. With a cumulative metric, the extra day changes the estimand and the variance. The additional data can, therefore, lead to less precision.

\section{When to use cumulative metrics}\label{sec:when_to_use_cum_metrics}
The previous sections have illustrated some undesirable properties of cumulative metrics. However, there are situations where there are clear benefits to using cumulative metrics, or where the problems are less pronounced.

\subsection{Cumulative metrics for error detection}
Online experimentation serves two primary purposes: to quantify the impact and to manage risks. The risk management part traditionally means ensuring that the false positive and false negative rates are bounded as intended. However, in online experimentation, an important part is also to quickly detect poor experiences to avoid exposing users to them. If a treatment in an experiment causes a worse experience, the regression must be detected quickly to abort the experiment. 

For example, suppose that we are interested in tracking the share of users in the experiment that experience a crash in their app as a guardrail metric. We have no particular interest in knowing when the crash happens, since the goal is simply to learn whether there is an increase in the share of users that experience a crash. Because of this, deciding on a single time frame to use with the windowed metric is challenging. If the window is too long, the experimenter must wait for the entire window length to pass to see any results on the metric. If the window is too short, users may not have had time to experience any crashes before they are measured.

\subsection{Temporary visitors lead to built-in windows}
Our analysis has focused on users with an implicit assumption that these users are recurring and identified using persistent identifiers. These identifiers enable the possibility to track users over time. In practice, that is not always possible. In some cases, users stay unauthenticated, which means they receive a new identifier when their old expires or when they clear that information from their device themselves. Another possibility is that the product they are a user of is an infrequently used product, so that its users visit the product, achieve their goal, and then do not return in the near future. 

When users are temporary visitors, cumulative metrics resemble windowed metrics more because of the induced windows. These induced windows are more fuzzy and probabilistic windows than what is used in the windowed metrics. For example, suppose that 50\% of users do not return after the end of the first week after exposure, and 95\% after the end of the second week after exposure. While not a strictly enforced window on what data on a user to include, almost no data exists for users beyond the second week after exposure. Beyond the second week, the cumulative metric will behave much like a windowed metric.

The implication of this is that the drawbacks we have raised about cumulative metrics are most severe for experiments on recurring users that can be tracked for longer periods of time. Experiments on users that are more temporary, for one reason or another, are less affected by the drawbacks of cumulative measurements.

\section{Discussion}

This paper has examined the properties of cumulative and windowed metrics in online controlled experiments, revealing implications for experimental design and analysis. In our investigation, we have highlighted the complexities and trade-offs present in metric selection for experimentation.

We have demonstrated that cumulative metrics, while widely used, yield estimands that vary with experiment runtime and the distribution of exposure. This dependence introduces a layer of complexity to result interpretation and makes comparisons across experiments potentially misleading. Experimenters using cumulative metrics may unknowingly be measuring different quantities in seemingly similar experiments, leading to inconsistent conclusions and decision-making.

Perhaps most strikingly, our analysis reveals that for cumulative metrics, the standard intuition that more data invariably increases statistical power does not always hold. This finding challenges fundamental approaches to experiment planning and sample size determination. It suggests that experimenters using cumulative metrics need to re-evaluate their strategies for determining experiment runtimes and sample size when using cumulative metrics, as extending an experiment may not always yield the expected improvements in precision or power.

Windowed metrics offer greater transparency and control over what is being measured at the cost of having to wait until the full window is observed before getting results. By explicitly defining the window of interest for treatment effects, experimenters can ensure that their analyses focus on the most relevant period for what they are interested in. This approach allows for more consistent comparisons across experiments. However, cumulative metrics are useful in many scenarios. For example, they can be particularly effective for error detection, where the exact timing of an issue is less important than whether it has occurred at all. Additionally, in contexts with primarily temporary visitors or where user tracking is limited, cumulative metrics may naturally behave more like windowed metrics due to time windows being an inherent part of user behavior.

No single type of metric is consistently superior across all experimental scenarios. The choice between cumulative and windowed metrics depends critically on the shape of the treatment effect curve as a function of time since exposure. Cumulative metrics can be more effective in detecting large initial effects and providing early insights, particularly useful for quick error detection or when immediate user responses are of primary interest. On the other hand, windowed metrics offer more stable and interpretable results over time. The choice between cumulative and windowed metrics should be made thoughtfully, in particular the primary goals of the experiment.

In our opinion, cumulative metrics and windowed metrics should be combined in online experiments. Cumulative or cumulative-windowed metrics can be used for tracking regressions to be able to stop experiments that are having a negative effect on users or systems. Windowed metrics can be used for product-decision making and impact quantification. Alternatively, cumulative-windowed metrics offer an appealing compromise that lets the experimenter view results and detect regressions faster, while having an estimator that converges to a clear estimand. 

\bibliography{main}
\bibliographystyle{abbrvnat}

\pagebreak
\onecolumn
\appendix

\subsection{Theorem 1}
\label{app:thm1}
\subsubsection{Cumulative metric}
\begin{align}
\E[Y(t)|W=1, E<t]-\E[Y(t)|W=0, E<t]=\frac{\int_0^t\Delta(t-e)\diff \cdf_{E}(e)}{\cdf_E(t)},
\end{align}
where 
\begin{align}
    \Delta(t-e)=\E[Y(t)|W=1, E=e]-\E[Y(t)|W=0, E=e].
\end{align}

\subsubsection{Windowed metric}
\begin{align}
\E[Y^{\textrm{win}}(t)|W=1, E<t-\nu]-\E[Y^{\textrm{win}}(t)|W=0, E<t-\nu]
    &=\frac{\int_0^{t-\nu} \Delta(\nu)f_{E}(e)de}{\cdf_E(t-\nu)}=\Delta(\nu).
\end{align}

\subsubsection{Cumulative windowed metric}
\begin{align}
\E[Y^\textrm{cwin}(t)|W=1, E<t]-\E[Y^\textrm{cwin}(t)|W=0, E<t]&=\frac{\int_0^{t-\nu} \Delta(\nu)\diff \cdf_{E}(e)}{\cdf_E(t)}+\frac{\int_{t-\nu}^t \Delta(t-e)\diff \cdf_E(e)}{\cdf_E(t)}\\
    &=\frac{\Delta(\nu)\cdf_E(t-\nu)+\int_{t-\nu}^t \Delta(t-e)f_{E}(e)de}{\cdf_E(t)}
\end{align}

\subsection{Theorem 2}
\label{app:thm2}
The decomposition of the difference between expected $Z$ statistics at time points $t$ and $t'$ ($t'>t$) can be derived as
\begin{align}
&V_{t'}^{-1/2}\tau_C^1(t)-V_t^{-1/2}\tau_C^1(t)\\
&=\frac{1}{V_{t'}^{1/2}\cdf_E(t')}\int_0^{t'}\Delta(t'-e)\diff \cdf_E(e)-\frac{1}{V_{t}^{1/2}\cdf_E(t)}\int_0^{t}\Delta(t-e)\diff \cdf_E(e)\\
&=\frac{1}{V_{t'}^{1/2}\cdf_E(t')}\int_0^{t}\Delta(t'-e)\diff \cdf_E(e)-\frac{1}{V_{t}^{1/2}\cdf_E(t)}\int_0^{t}\Delta(t-e)\diff \cdf_E(e)
+\frac{1}{V_{t'}^{1/2}\cdf_E(t')}\int_t^{t'}\Delta(t'-e)\diff \cdf_E(e)\\
&=\int_0^t\left[\frac{1}{V_{t'}^{1/2}\cdf_E(t')}\Delta(t'-e)
-\frac{1}{V_{t}^{1/2}\cdf_E(t)}\Delta(t-e)\right]\diff \cdf_E(e)+\frac{1}{V_{t'}^{1/2}\cdf_E(t')}\int_t^{t'}\Delta(t'-e)\diff \cdf_E(e)\\
&=\int_0^t\left[\frac{1}{V_{t'}^{1/2}\cdf_E(t')}\left(\Delta(t-e)+\int_{t-e}^{t'-e}\diff \Delta(s)\right)-\frac{1}{V_{t}^{1/2}\cdf_E(t)}\Delta(t-e)\right]\diff \cdf_E(e)\\
&+\frac{1}{V_{t'}^{1/2}\cdf_E(t')}\int_t^{t'}\Delta(t'-e)\diff \cdf_E(e)\\
&=\left[\frac{1}{V_{t'}^{1/2}\cdf_E(t')}-\frac{1}{V_{t}^{1/2}\cdf_E(t)}\right]\int_0^t\Delta(t-e)\diff \cdf_E(e)\\
&+\frac{1}{V_{t'}^{1/2}\cdf_E(t')}\int_0^t\int_{t-e}^{t'-e}\diff \Delta(s)\diff \cdf_E(e)+\frac{1}{V_{t'}^{1/2}\cdf_E(t')}\int_t^{t'}\Delta(t'-e)\diff \cdf_E(e)
\end{align}

\subsection{Example 2}
\label{app:ex2}
The variance of the treatment group estimator is
\begin{align}
    \Var\left(\hat{\mu}_t^{(1)}|\mathcal{R}_t\right)
    &=\Var\left(\frac{1}{N_t^{(1)}}\sum_{i=1}^NR_i(t)W_i Y_i(t)|\mathcal{R}_t\right)\\    
    &=\left(\frac{1}{N_t^{(1)}}\right)^2\Var\left(\sum_{i:R_i(t)=1\land W_i=1} Y_i(t)|\mathcal{R}_t\right)\\  
    &=\left(\frac{1}{N_t^{(1)}}\right)^2\sum_{i:R_i(t)=1\land W_i=1} \Var\left(Y_i(t)|\mathcal{R}_t\right)\\    &=\left(\frac{1}{N_t^{(1)}}\right) \Var\left(Y(t)|E<t\right),
\end{align}
with a similar result for the control group.

Decompose the variance using the law of total variance
\begin{align}
\Var(Y(t)|E<t)=\E\left[\Var(Y(t)|E)|E<t\right]+\Var\left[\E(Y(t)|E)|E<t\right],
\end{align}
where
\begin{align}
\E\left[\Var(Y(t)|E)|E<t\right]&=\frac{\int_0^t \Var[Y(t)|E=e]\diff \cdf_E(e)}{\cdf_E(t)}=\frac{\sigma^2\int_0^t (t-e)\diff \cdf_E(e)}{\cdf_E(t)}=\sigma^2\left(t-\E(E|E<t)\right)
\end{align}
and
\begin{align}
\Var(\E(Y(t)|E)|E<t)&=\Var\left[\mu + \Delta(t-E)|E<t\right]=\Var\left[\Delta(t-E)|E<t\right].
\end{align}

In the example, all users are exposed at time 0 or time 7. For $0\leq t<7$, we know that $\Pr(E=0|E<t)=1$. This means there is no variation of the expected treatment effect, because each exposed user has $\Delta(t-E)=\Delta(t)$. Similarly, for $t\geq 14$ it holds that $\Delta(t-7)=\Delta(t-14)=\Delta(7)$ and there is again no variation.

For $7<t<14$,
\begin{align}
\Var[\Delta(t-E)|E<t]&=(0.5\times (\Delta(7))^2+0.5\times (\Delta(t-7))^2)-(0.5\times \Delta(7)+0.5\times \Delta(t-7))^2\\
&=0.5\left[\left(\Delta(7)^2+\Delta(t-7)^2\right)-0.5\left(\Delta(7)+\Delta(t-7)\right)^2\right].
\end{align}
Because $\Delta(7)=7c$ and $\Delta(t-7)=(t-7)c$, we obtain
\begin{align}
\Var[\Delta(t-E)|E<t]&=0.5\left[c^2\left(98+t^2-14t\right)-0.5t^2c^2\right]\\
&=\left[c^2\left(49+0.25t^2-7t\right)\right]\\
&=\left[0.25c^2\left(196+t^2-28t\right)\right]\\
&=0.25c^2\left(t-14\right)^2.
\end{align}

For $t\geq 14$, the variance is zero again because all have the same expected treatment effect.

This means that
\begin{align}
   \Var(Y(t)|E<t)=\begin{cases}
    \sigma^2t, &\quad t<7,\\
    \sigma^2\left(t-3.5\right)+0.25c^2\left(t-14\right)^2, &\quad 7\leq t<14,\\
    \sigma^2\left(t-3.5\right), &\quad t\geq 14.
    \end{cases}
\end{align}

The numerator of the $Z$ statistic is the estimand, which here is 
\begin{align}
    \E(\hat{\mu}_t^{(1)}-\hat{\mu}_t^{(0)})=\begin{cases}
        tc, &\quad t< 7\\
        0.5 tc, &\quad 7\leq t<14\\
        7c, &\quad t\geq 14
    \end{cases}
\end{align}

The expected $Z$ statistics are therefore
\begin{align}
    \E(Z_t)=\begin{cases}
        \frac{125c}{\sigma^2}, &\quad t< 7\\
        \frac{125tc}{\sigma^2\left(t-3.5\right)+0.25c^2\left(t-14\right)^2}, &\quad 7\leq t < 14\\
        \frac{250\times 7c}{\sigma^2(t-3.5)}, &\quad t\geq 14.
    \end{cases}
\end{align}
\end{document}